\documentclass[12pt]{article}
\makeatletter
\newcommand{\singlespacing}{\let\CS=\@currsize\renewcommand{\baselinestretch}
{1.0}\tiny\CS}
\newcommand{\doublespacing}{\let\CS=\@currsize\renewcommand{\baselinestretch}
{1.5}\tiny\CS}

\begin{document}
\title{Edge Current of FQHE and Aharanov-Bhom Type Phase}
\author{Dipti Banerjee \\
\singlespacing Department of Physics \\
Rishi Bankim Chandra College \\
Naihati,24-Parganas(N)\\
Pin-743 165, West Bengal\\
INDIA}
\date{}
\newcommand{\bd}{\begin{document}}
\newcommand{\ed}{\end{document}}
\newcommand{\bc}{\begin{center}}
\newcommand{\ec}{\end{center}}
\newcommand{\bfr}{\begin{flushright}}
\newcommand{\efr}{\end{flushright}}
\newcommand{\vs}{\vspace}
\newcommand{\hs}{\hspace}
\newcommand{\beq}{\begin{equation}}
\newcommand{\eeq}{\end{equation}}
\newcommand{\lb}{\linebreak}
\newcommand{\pb}{\pagebreak}
\newcommand{\mb}{\makebox}
\newcommand{\fb}{\framebox}
\newcommand{\mc}{\multicolumn}
\newcommand{\ben}{\begin{enumerate}}
\newcommand{\een}{\end{enumerate}}
\newcommand{\bit}{\begin{itemize}}
\newcommand{\eit}{\end{itemize}}
\newcommand{\ol}{\overline}
\newcommand{\un}{\underline}
\newcommand{\lefq}{\lefteqn}
\newcommand{\ba}{\begin{array}}
\newcommand{\ea}{\end{array}}
\newcommand{\beqa}{\begin{eqnarray}}
\newcommand{\eeqa}{\end{eqnarray}}
\newcommand{\beqas}{\begin{eqnarray*}}
\newcommand{\eeqas}{\end{eqnarray*}}
\newcommand{\bfg}{\begin{figure}}
\newcommand{\efg}{\end{figure}}
\newcommand{\bds}{\begin{displaymath}}
\newcommand{\eds}{\end{displaymath}}
\newcommand{\btb}{\begin{tabbing}}
\newcommand{\etb}{\end{tabbing}}
\newcommand{\para}{\parallel}
\newcommand{\pad}{\partial}
\newcommand{\nn}{\nonumber}
\newcommand{\la}{\leftarrow}
\newcommand{\ra}{\rightarrow}
\newcommand{\lgla}{\longleftarrow}
\newcommand{\lgra}{\longrightarrow}
\newcommand{\La}{\Leftarrow}
\newcommand{\Ra}{\Rightarrow}
\newcommand{\Lra}{\Leftrightarrow}
\newcommand{\Lgla}{\Longleftarrow}
\newcommand{\Lgra}{\Longrightarrow}
\newcommand{\bm}{\boldmath}
\newcommand{\lan}{\langle}
\newcommand{\ran}{\rangle}
\renewcommand{\a}{\alpha}
\renewcommand{\b}{\beta}
\newcommand{\g}{\gamma}
\newcommand{\G}{\Gamma}
\renewcommand{\d}{\delta}
\newcommand{\eps}{\epsilon}
\newcommand{\Th}{\Theta}
\newcommand{\s}{\sigma}
\newcommand{\lam}{\lambda}
\newcommand{\D}{\Delta}
\newcommand{\vare}{\varepsilon}
\newcommand{\pr}{\prime}
\newcommand{\ro}{\rho}
\newcommand{\nab}{\nabla}
\newcommand{\m}{\mu}
\newcommand{\n}{\nu}
\newcommand{\Sg}{\Sigma}
\newcommand{\p}{\pi}
\newcommand{\R}{I\!\!R}
\newcommand{\om}{\omega}
\newcommand{\Om}{\Omega}
\newcommand{\ze}{\zeta}
\newcommand{\vart}{\vartheta}
\newcommand{\tri}{\triangle}
\newcommand{\f}{\frac}
\newcommand{\iny}{\infty}
\newcommand{\pro}{\propto}

\maketitle

\abstract When two non-identical quasi-particles in the Hall fluid
encircle each other, relative AB type phase developes.As the
quasi-particles advance towards the edge in a similar circular
way, the developed current should have connection with this AB
type phase through the {\it Shift} quantum number or Berry's
topological phase. We have pointed out the role of relative AB
type statistical phase in the development of edge current.In
fact,the physics of the current flow in FQHE is sketched here from
the topological point of phase transformation.

 \vspace{3cm}
email:deepbancu@hotmail.com

\pagebreak

\section{\bf Introduction}

 A fundamental feature of the microscopic theory is the existence
 of a one-to-one correspondence between
  the quasi particle states of the bulk and the primary
fields that build the spectrum of the edge states.Microscopically
the bulk states can be constructed by implementing the idea of
flux-attachment by coupling suitable set of Chern-Simon gauge
fields[1]. The effective action in the field theory is a Chern-Simon gauge theory
 that fits the K-matrix classification
[2] reflecting the underlying hierachical construction of FQH
states. The composite picture of the edge states have number of
branches that is encoded in the rank of the K-matrix.
Wen has shown that edge excitations in FQH states provide an important
probe to detect the topological orders in the bulk FQH states.

At the edge, the electrostatic potential varies very slowly and adiabatic
conductance takes place between series of alternating compressible and incompressible
strips forming channels at the edge. The QH liquid for $\nu=1/m$ contains only one component
of incompressible state that lead to one branch. A generic QH state with $\nu \neq 1/m$
contains many branches of edge excitations.
The concept of edge channels is extended from the integer to the fractional quantum Hall
effect, and the contribution of an adiabatically transmitted edge channel to the conductance
is calculated from the point of view of interacting-electron picture [3].
 For a sufficiently small $\Delta\mu$ (chemical potential)
the current carried by quasi-particles in a compressible region at the edge depends on the                                                        on the
difference of the electron filling factors in the two adjacent incompressible
regions[4].

It is well known that these quasi-particles in FQHE are not fundamental particles and
obey fractional statistics in two dimensions.
Any fractional statistics objects are collective particles of a nontrivial
condensed matter state. The fractional statistics as pointed out
by Leinaas and Myrheim [5] relies on the property that when
particles with infinitely strong short range repulsion are
confined in two dimensions, paths with different winding numbers
are topologically distinct and cannot be deformed into one
another. The particles [6] are said to have statistics $\theta$
producing a phase factor $e^{i\pi\theta}=(-1)^\theta$ when exchange of particles takes
place over a half loop. Non integral values of $\theta$ imply fractional statistics.

 It is believed [7] that fractional statistics are the consequences of
incompressibility at a fractional filling and may possible be observable
in an experiment specially designed for this purpose.
The fractional statistics can be derived heuristically in the
Composite Fermion theory [8] when one CF goes around another
encircling an area A, the total phase associated with this path is
given by
$$\Phi^*= - 2\pi(BA/\phi_0 - 2pN_{enc})$$
where $N_{enc}$ is the number of composite fermions inside the
loop. The first term on the right-hand side is the usual AB phase
and the second term is the contribution from the vortices bound to
composite fermions indicating  that each enclosed CF effectively
reduces the flux by $2p$ flux quanta.
 Wilczek shed considerable light on fractional statistics of quasi particles
[9].In two dimensions these quasi-particles are similar as vortex
attached at the point particle. If a vortex is dragged
adiabatically around a closed loop,the system will acquire an
extra non dynamical phase which can be gauged away resulting
continuously from one incompressible liquid state to another at
different filling fraction.
The concept of fractional statistics has been reformulated by
Haldane [10] as a generalization of Pauli exclusion principle and a
definition independent of the dimension of space. In the FQHE, the
Pauli like definition of statistics can be introduced in the quasi
particles which are flux carrying charged bosons in the lowest
Landau level.If an object carrying flux $\phi_\alpha$ and charge
$q_\alpha$ orbits around another object carrying flux $\phi_\beta$
and charge $q_\beta$ the relative statistical phase
$\theta_{\alpha\beta}$ becomes
\begin{equation}
exp(i\theta_{\alpha\beta})= exp {\pm i\pi(g_{\alpha\beta}+g_{\beta\alpha}})
\end{equation}
 where $$g_{\alpha\beta}=-q_{\alpha} \phi_{\beta}$$
With these view we have recently shown [11] that due to
interchange of two nonidentical composite fermions residing in two
consecutive Landau levels,the relative AB type phase developed and
{\it shift} quantum number can be visualized through it. In
another issue [12] we have showed that this {\it shift} vector
have connection with the change of edge current through the
difference of filling factor $\nu_n-\nu_{n-1}$ between two
consecutive incompressible states of the edge. In fact a more
meaningful idea of the {\it shift} [13] has been given through
deviations of Berry's topological phase of composite particles in
hierarchies.

Motivated by the recent two works, one which highlights the
importance of non equilibrium noise measurement through
statistical phase [14] and other is FQHE qubits in connection with
topological quantum computation [15], we realize that in the
transport process of edge current, the relative AB type phase
should play a major role. Hence we here want to find the origin of
the edge current developed through the statistical interaction of
composite particles in the AB type topological phase.

\section{\bf Topological Aspect of Hall fluid, Berry Phase and Shift quantum number}

    In the Hall fluid the statistical interaction takes the most significant role. Being
    long ranged it is treated non-perturbatively.
    A non-dynamical gauge field $A_\m$ is associated
    with the flux which in $2+1$ dimension is the very cause of the
    appearance of Chern-Simon term in the Lagrangian.
    \beq
    L_{CS} = \f{\m}{2}\eps^{\m\n\lambda} A_\m \pad_\n A_\lambda
    \eeq
    This licenses a conservation of topological current $J_\m$ which
    include a topological invariant term in the $(2+1)$ dimension [9]
    \beq H = \f{\theta}{2 \p} \int d^3 x A_\m J^\m \eeq in the action.
    In fact it is the Hopf invariant describing basic maps of $S^3$ to
    $S^2$. If $\rho$ denotes a four dimensional index then we find
    \beq
    \pad_\rho \eps^{\rho \m \n \lambda} A_\m F_{\m\n} = \f{1}{2} \eps^{\rho \m \n
    \lambda} F_{\rho \m} F_{\n\lambda}
    \eeq
    which connects the Hopf invariant with chiral anomaly. This Hopf term plays a
    role somewhat similar to the role played by the Wess-Zumino interaction in
    connection with $3+1$ dim. Skyrmion term.

    There is an analogous statistical interaction in (3+1) dimensions given by
    Haldane [16] considering the 2D Hall surface as a
    boundary surface of a 3D sphere, having radius R in a radial (monopole) magnetic
    field $B=\hbar S/eR^2 (>0)$. This $2S = N_\phi$ is an integer which defines
    the total number of magnetic flux through the surface. For the parent state $\n
    = 1/m$ the total flux is $S = \f{1}{2} m (N - 1)$. The field strength $S$ in
    the first level hierarchy is
    $$S (N, m \pm p) = \f{1}{2}m(N - 1)\pm \f{1}{2} (\f{N}{p} + 1)$$
    which is formed when $p$ ($p$= even integer) excitations are added in the
    parent state $\n = \f{1}{m}$. These show that the filling factors  for hierarchical
    state satisfy a slight complicated relation.
 $$2S = N \phi = \n^{-1} N - \cal{S}$$
In the language of Wen and Zee [1,2] this $\cal S$ is the {\it shift}, a topological
quantum number which is developed due to the coupling between the orbital spin
 and the curvature of the space having spin $s=\frac{1}{2}
 K_{II}$. On a sphere,the {\it shift} for a hierarchical state is
 given by
\beq
{\cal S} =\frac{1}{\nu} \sum_{IJ} {(K^{-1})}_{IJ} K_{JJ}
\eeq

  For a $\nu=\frac{1}{m}$ parent
state this {\it shift} is simply  ${\cal S}=2(n-1)+m$ having orbital spin
$s=n-1+\frac{m}{2}$ that is associated with the orbital angular momentum in cyclotron motion.
  In the effective theory, this introduction of {\it shift} leads to a modification
   of the Lagrangian in equation (2) as follows
\begin{equation}
{\cal L}=1/4\pi(K B \epsilon \partial B + 2Ae\epsilon \partial B
  + 2Cs\epsilon \partial B)
\end{equation}
where the second term is the electromagnetic coupling and the
third one is the coupling to the curvature of space.

The appearance of {\it shift}
 in the hierarchies of FQHE is nontrivial [13].
The quasiparticles in these levels are formed when additional
fluxes are attached with quantized particles.
In fact the quantization of Hall particles is the indication of Quantum Hall
  effect involving gauge theoretic extension of coordinate by $C_\mu
\in SL(2C)$ which is visualized through the field strength
$\tilde{F_{\mu\nu}}$. Apart from the internal extension, the
external strong magnetic field induces gauge extensions $B_\mu\in
SL(2C)$ through the gauge field $F_{\mu\nu}$.In the language of
differential geometry these two gauges act as two fibres at each
particle points of the base space $S^2$. The effective theory of
the Hall fluid (Abelian) can be accurately presented if not only
the two vortices but also their interactions are taken into
account. In the light of Haldane[16], we consider the Hall surface
on 3D sphere and in the presence of strong external magnetic
field, the chiral symmetry breaking of composite fermion
associated with internal and external gauge fields $F_{\mu\nu}$
and $\tilde{F_{\mu\nu}}$ are represented by
$$
F_{\mu\nu}=\partial_\mu B_\mu - \partial_\nu B_\nu + [B_\mu,
B_\nu]$$
\begin{equation}
\tilde{F_{\mu\nu}}=\partial_\mu C_\mu - \partial_\nu C_\nu +
[C_\mu, C_\nu]
\end{equation}
In particular the $\theta$ term in the Lagrangian leads to vortex line
and the corresponding gauge field acts like a magnetic field.
 The topological Lagrangian of Hall fluid  can be described by
 the added Chern-Simon terms in the Lagrangian through the anomaly [13]
\begin{equation}
{\cal L}= -\frac{\theta}{16\pi^2} Tr^* F_{\mu\nu}F_{\mu\nu}
-\frac{\theta^\prime}{16\pi^2} Tr^* F_{\mu\nu}\tilde{F}_{\mu\nu}-
\frac{\theta^{\prime\prime}}{16\pi^2}
Tr^*\tilde{F}_{\mu\nu}\tilde{F}_{\mu\nu}
\end{equation}
Here every term corresponds to a total divergence of a topological
quantity, known as Chern-Simons secondary characteristics class
defined by
\begin{equation}
{\Omega^\mu}_e =-\frac{1}{16\pi^2 \epsilon_{\mu\nu\alpha\beta}}
Tr[B_\nu F_{\alpha\beta}-2/3(B_\nu B_\alpha B_\beta)]
\end{equation}

\begin{equation}
{\tilde{\Omega}^\mu} =-\frac{1}{16\pi^2
\epsilon_{\mu\nu\alpha\beta}} Tr[C_\nu F_{\alpha\beta}-2/3(C_\nu
B_\alpha B_\beta)]
\end{equation}

\begin{equation}
{\Omega^\mu}_i=-\frac{1}{16\pi^2 \epsilon_{\mu\nu\alpha\beta}}
Tr[C_\nu \tilde{F}_{\alpha \beta} - 2/3(C_\nu C_\alpha C_\beta)]
\end{equation}
Assuming a particular choice of coupling $\theta=\theta'=\theta"$
in the Lagrangian the topological part of the action in (3+1)
dimension become
\begin{equation}
W_{\theta}=2(\mu_e + \mu_i + \tilde{\mu})\theta
\end{equation}
where $\m_e$, $\m_i$ and $\tilde \m$ are the corresponding
magnetic charges which are connected with the respective charges
through the Dirac quantization condition and Pontryagin density.
\beq 2 \m=q=\int \pad_\m \Omega_\m  d^4 x \eeq

 It gives rise the topological phase of Berry on the parallel
transport over a closed path of a Hall hierarchy state.
\begin{equation}
\phi_B = \pi W_\theta = 2\pi{\tilde{\mu}}_{eff}\theta = 2\pi(\mu_{eff}+\tilde{\mu})\theta
= 2\pi(\mu_e+\mu_i+\tilde{\mu})\theta
\end{equation}
Here the first term is associated with Berry phase factor of Hall
particle due to external magnetic field $\mu_e$. The second term
gives rise to the inherent Berry phase factor $\mu_i$ associated
with the chiral anomaly of a free electron (in absence of an
external magnetic field) and the third one effectively relates the
coupling of the external field with the internal one which give
rise the phase factor $\tilde{\mu}$. This $\mu_{eff}$ actually
visualizes the filling factor through the relation
$\nu=\frac{n}{2\mu_{eff}}$. where $n$ denotes the $nth$ Landau
level. In fact this $\mu_{eff}$ satisfies the Dirac quantization
condition
\begin{equation}
e^{\prime}\mu_{eff}=\frac{n}{2}
\end{equation}
showing that each quasi particle in the $n^{th}$ Landau level
having charge $e^{\prime}$ behaves as a composite fermion. It will
behave as fermion in the ground state following the Dirac
condition
\begin{equation}
\tilde{e\mu}=\pm1/2
\end{equation}
which follows the equation
\begin{equation}
\tilde{e}(\mu_{eff} - \frac{n\pm 1}{2})=\pm\frac{1}{2}
\end{equation}
This implies that $(n\pm1)/2$ is the magnetic strength $\mu$ of
the added quanta whose removal makes the composite fermion in the
higher Landau levels to behave fermion in the ground state. Here
for $\mu=\pm 1/2,\pm 3/2,...$ the quanta behave like fermion and
$\mu=\pm1,\pm2,....$ it shows bosonic behavior.
We have found this
change of magnetic charge as $\tilde{\mu}$ that can be visualized
through {\it shift} ${\cal S}$ by the relation
\begin{equation}
2\tilde{\mu}={\cal S}= 2\mu_{eff}-(n\pm 1)=\frac{n}{\nu}-(n\pm 1)
\end{equation}
where $n=1,2,3..$ denotes the hierarchy levels.

Our picture shows that a motion of composite particle in the Hall
fluid moving in a circular orbit will be quantized through
 its acquirance of the Berry's topological phase.
\begin{equation}
\phi_B = \pi\theta(2\mu_{eff}+\cal S)
\end{equation}
Conceptually the appearance of this {\it shift} quantum number
$\cal S$, in the topological phase of quasi-particle is obvious,
since the coupling between the two gauges(that act as fibres) with
the curvature is prominent during parallel transport over a closed
path.

With this view we have found [11]the role of {\it shift} in the
relative AB type phase as the composite fermion and the additional
quanta encircles each other for producing fermions in the lowest
Landau level. In addition, we have shown elsewhere [12] in the
context of edge current flowing through the compressible level
 that this {\it shift} can be related with the difference of Landau filling factor
  between two consecutive incompressible levels.
  \begin{equation}
  \frac{\tilde{e}}{2\mu_eff}{\cal S}=(\nu_n-\nu_{n-1})
  \end{equation}
  For a sufficiently small chemical potential $\Delta\mu$,
   the change of current in a compressible strip is,
  \begin{equation}
  \Delta I= \frac{e}{h} \Delta\mu \Delta\nu =  \frac{e}{h}
  \Delta\mu( \nu_n-\nu_{n-1})
  \end{equation}
  that can be expressed in terms of {\it shift}
  \begin{equation}
  \Delta I=-\frac{e}{h}\Delta\mu \frac{\tilde{e}}{2\mu_eff}{\cal S}
  \end{equation}
Now we will proceed to find the role of A-B type statistical phase
in the development of edge current as the composite particles
advance on the Hall surface.

\begin{center}
\section{\bf The Edge current in A-B Type Phase}
\end{center}

The concept of  edge channels for IQHE and FQHE in combination with the adiabatic
transport of quasi-particle is successful in explaining the anomalous dependence
 of Hall conductance. Edge channels are defined with the
 correspondence of bulk landau level.On approaching the boundary
 of the 2DEG a Landau level which in the bulk lies below the Fermi
 level rises in energy because of the presence of confining
 potential. The intersection between the nth Landau level and the
 Fermi level defines the location of the nth edge channel for
 filling factor in the nth hierarchy. In general the current
 injected into $pth$ edge channel is [3]
 \begin{equation}
 I_p = \frac{e}{h}\Delta\mu(\nu_p - \nu_{p-1})
 \end{equation}
 where the current $I_p$ in a compressible band is in between two incompressible bands
 of filling factors $\nu_p$ and $\nu_{p-1}$.

The tunnelling current $I$ through the wire is $I(V)\propto V^\alpha$,
where the exponent $\alpha$ is determined by the
scaling dimension of the tunnelling operator. Lopaz and Fradkin
[17] pointed out recently that in case of tunnelling of electrons
from Fermi liquid into a hierarchical FQH state, the tunnelling
exponent is $\alpha= 1/\nu$. The physics behind the tunnelling in
the edge has been focussed on the charge and neutral modes that are
propagating with different velocities. This latter one has been
identified as topological  mode  which is responsible for
Fermi statistics.Representing the respective charge mode and
topological mode by $\phi_c$ and $\phi_T$, a general edge operator
is
\begin{equation}
\psi (x) = exp i ( \alpha_c \phi_c +\alpha_T \phi_T +\Sigma \alpha_T \phi_T )
\end{equation}
The authors have shown that charge $Q$ depend only on $\alpha_c$
 but the statistics $\theta$ of these excitations is connected
 with both $\alpha_c$ and $\alpha_T$.

In a recent communication Zulicke and Mac Donald et.al.[18] addressed
the chiral phase field $\phi_n (\theta)$ as a superposition of
edge-density fluctuations.
\begin{equation}
\phi_n (\theta) =\frac{1}{\sqrt{\nu}}\phi^c (\theta) + \xi_n \phi^n (\theta)
\end{equation}
where $\phi^c$ is the phase field of the charged edge-magneto plasmon mode
which corresponds to fluctuations in the total edge-charge density and $\phi^n$
is its orthogonal complement known as neutral mode.The authors have expressed
these two modes in terms of the parent state $\phi_0$ and daughter state
$\phi_{2p+1}$ with the respective filling factors $\nu_0 = \frac{1}{2p+1}$ and
$ \nu_i = \frac{1}{(2p+1)(4p+1)}$, that comprise
the $\nu = \frac{2}{4p+1}$ QH state.
The addition of electrons to the edge with concomitant change of
$2p+1-n$ flux quanta is viewed as adding the electron to the outer
edge and transferring at the same location $n$ fractionally charged
quasiparticles from the outer QH droplet to the inner one.

We realize from the above works that both the charge and neutral modes are transferred
from the inner edge to outer edge leading to flow of current and change of statistics.
We are now interested whether this current and statistics
are interrelated during the course of transfer towards the edge.
Inspired by the works on topological transformation of
current in the FQHE system  in connection with Quantum Computation [15] through
fractional statistics, we now proceed to evaluate the role of
AB type Statistical phase in edge current flow two ways.  \\
1. In a particular edge the  composite particles in the consecutive branch (Landau level)
 encircles during transformation.\\
2. From inner edge the composite particles transform  to the outer edge picking up integral multiple
of flux from the bulk. \\

It is now known that composite particles in FQHE are the composite of fluxes attached with
charged particle.When an electron  is attached with a magnetic flux, its statistics  changes
and it is transformed into a boson.These bosons condense to form cluster which is
coupled with the residual fermion or boson (composed of two fermions). Indeed the residual
boson or fermion will undergo a statistical interaction tied to a geometric Berry phase
effect that winds the phase of the particles as it encircles the vortices. Also we observe
that the attachment of vortices to electrons in a cluster will make the fluid an incompressible
one. Indeed as two vortices cannot be brought very close to each other, there will be a hard core
repulsion in the system which accounts for the incompressibility of the Quantum Hall fluid.
In fact the Hall particles are quantized by acquiring Berry's topological phase
as discussed in sec.-2. As the quasi-particles encircles another in their way of topological
transport, the Aharanov-Bhom type statistical phase is developed.

At first, we concentrate on one edge of a QH system where we find current in the
compressible band (eqn.-23) depends on the filling factors of the consecutive
 incompressible Landau levels $n^{th}$ and $(n-1)^{th}$ respectively.
  During this movement of quasiparticles, the charge
dressed with flux advance following circular path.As one encircles another relative
 AB type phase developed [10]
\begin{equation}
\phi_s=exp \pm{\frac{i\pi}{2}}(q_n \mu_{n-1}+ q_{n-1}\mu_n)
\end{equation}
where $q_n, q_{n-1}$ are the respective charges and $\mu_n, \mu_{n-1}$ are the
corresponding magnetic strength of the flux attached.These composite particles
follow Dirac Quantization condition $q_n \mu_n=n/2$ having respective filling factors
$\nu_n=\frac{n}{2\mu_n}$ and $\nu_{n-1}=\frac{n-1}{2\mu_{n-1}}$. Now the intertwining
of these Composite particles against each other results
 \begin{equation}
 \phi_s=exp{\pm{\frac{i\pi}{2}}(\nu_n \mu_{n-1}+ \nu_{n-1} \mu_n)}
 \end{equation}
 \begin{equation}
 \phi_s=exp{\pm{\frac{i\pi}{2}(\frac{{\nu_n}^2(n-1)+{\nu_{n-1}}^2
 n}{2\nu_{n-1}\nu_n}}})
 \end{equation}
 After a few mathematical steps and using eqn.-23 we have
 \begin{equation}
 \phi_s = exp \pm{\frac{i\pi}{2}}(\frac{nK^2{I}^2}{2\nu_{n-1}\nu_n}+ 2n-\frac{\nu_n}{2\nu_{n-1}})
 \end{equation}
 where $K=\frac{e}{h}\Delta\mu$.\\
This implies that edge current or its change can be realized through the acquirance
of AB type statistical phase whenever two quasi-particles in the consecutive
Landau level encircle each other.In other words the noise in
the current flow is the very cause of these type of phase factor.

In the second case,we consider the edge tunnelling through
the bulk of FQHE. We assume the transfer of the composite
particle from the inner edge in the $n^th$
Landau level having filling factor $\nu_n$ picking up even integral
$(2m)$ of flux $\nu_1$ through the bulk of QH system and forming a new
composite particle in the $(n+1)^th$ Landau level of the outer edge.
The filling factor of the effective particle becomes
$\nu_{eff}=\frac{n+1}{\mu_{eff}}$.In the light of Haldane [16] and Jain [19],
we consider that the monopole strength  $\mu_{eff}$ of
the state ${\Phi_{1}}^{2m}\Phi_n$ can be obtained by noting that the
product of two monopole harmonics $\mu_1$ and $\mu_{n}$ gives a
monopole harmonic at $\mu_1+\mu_{n}$ i,e monopole strength add as follows
\begin{equation}
\mu_{eff}= 2m(\frac{N-1}{2}) + \frac{N-n^2}{2n}
\end{equation}
which can be considered as
\begin{equation}
\mu_{eff}=2m\mu_1+\mu_n.
\end{equation}
  Here statistical interaction takes place between the
  composite particle of the inner edge and outer edge which
  result current propagation. We further assume that path of the
  particles do not intersect each other. Encircling one type of
  fluxes around another in the consecutive Landau level relative AB
  type phase produces that is the very cause of  edge current
  flow.

 Following Haldane [10], we consider encircling of the composite particle
 in the inner edge having flux $\mu_n$ with charge $q_n$ that is
 equivalent to the filling factor $\nu_n=\frac{n}{2\mu_n}$, around
 the composite particle in the outer edge having corresponding flux and
 charge respectively $\mu_{eff}$ and
 $$q_{eff} = \frac{n+1}{2\mu_{eff}}= \frac{n+1}{2m\mu_1+ \mu_n}$$.
This generate the relative AB type phase developed
by their fluxes and charges as
\begin{equation}
\phi_s=exp \pm{\frac{i\pi}{2}}(q_n\mu_{eff}+q_{eff}\mu_n)
\end{equation}
Since the quasiparticles satisfy the Dirac quantization relation we can
write the above equation as
\begin{equation}
\phi_s=exp\pm{\frac{i\pi}{2}}(\nu_n \mu_{eff}+ \nu_{eff}\mu_n)
\end{equation}
After a few mathematical steps we found similar equation (as eq.-29) of
 the relative AB type statistical phase
\begin{equation}
\phi_s=exp\pm{\frac{i\pi}{2}}(\frac{nK^2 I^2}{2\nu_{eff}\nu_n}
+ 2n - \frac{\nu_n}{2\nu_{eff}})
\end{equation}
developed due to the transfer of a composite particle from
inner edge to the  outer edge through the bulk carrying the
integral multiple of flux alongwith.
We see that we get identical result in our two different approaches
of edge current flow.

From  another point of view using equation 31 we have
 \begin{equation}
 \phi_s=exp\pm{\frac{i\pi}{2}}(\frac{n}{\mu_n}(2m\mu_1+\mu_n)+\frac{(n+1)\mu_n}{(2m\mu_1+\mu_n)}
 \end{equation}
\begin{equation}
\phi_s \cong  exp\pm{\frac{i\pi}{2}}[\frac{n}{2}(2m
\frac{\mu_1}{\mu_n}+1)+ \frac{n+1}{2} (1- 2m\frac{\mu_1}{\mu_n})]
\end{equation}
 \begin{equation}
\phi_s \cong exp\pm{\frac{i\pi}{2}}[(n+\frac{1}{2})- m
\frac{\mu_1}{\mu_n}]
\end{equation}
 With our previous knowledge we see that this phase factor has
 relationship with {\it shift} quantum number. Above all we can
 comment from the above equation that irrespective of $\mu_1$ and $\mu_n$
being fermionic or bosonic flux, the phase factor
depends upon the number of particles-$N$,the Landau level-$n$,
and the odd integer-$m$ that is the inverse of parent filling factor
$\nu=1/m$ by the following expression.
\begin{equation}
\phi_s \cong exp\pm{\frac{i\pi}{2}}[(n+\frac{1}{2}) - \frac{m
n(N-1)}{(N-n^2)}]
\end{equation}
 Recently
it has been found that fractional statistics play an important role
in topological transformation in connection with Quantum computation [15].
Also Kane [14] showed that statistical
phase by combining AB effect can be used in noise measurement.
The result we find fully support these views.
The current obtained from the contacts of the Hall edge considering the
effect of bulk also can be visualized through AB type quantum phase.

\begin{center}
{\bf Discussions}
\end{center}
Quantization of Hall particles in the hierarchical states ensure
the acquirance of Berry's topological phase that visualize the
resultant chirality of the hierarchical state. Interchange of two
identical quasi particles develop statistical phase. Whereas two
dissimilar quasi particles on encircling each other produces
relative AB type phase. In this paper, we shed light on the latter phase
for its role in the edge current flow. At the edge, nonzero current
appears in the hierarchies which originate from the non-vanishing anomaly
in terms of the deviation of the topological phase through the difference of filling factors.
The quasi particles responsible for current flow, encircle one around other
in course of their advancement towards edge of Hall surface. As a result
relative AB type statistical phase evolved for the intertwining of fluxes around the charges
of the quasi particles in the following two cases.\\
1. In a particular edge the quasi-particles in the consecutive branch (incompressible
Landau level) encircles during transformation.\\
2. From inner edge the quasi-particles flow to the outer edge picking up integral multiple
of flux from the bulk of the Hall system.\\
 We find that in both the cases AB type statistical phase
is directly connected with edge current. And the second case combine the
physics of the edge and the bulk for the current flow.
 Hence classical current can be visualized through quantum phase.
In future these findings will help us to work for {\it Quantum Computation with FQHE qubits}[13]
and {\it Spin propagation in the Spintronics} devices [20].

{\bf Acknowledgement}\\
 I like to express my gratitude to all the
authors in my references.

\begin{center}                                                                                                                                                      1
\section {\bf References}
\end{center}
\singlespacing

\noindent [1] X.G.Wen  and A.Zee, Phys. Rev.-{\bf B46},2290 (1992)\\
\noindent[2] X.G.Wen and A.Zee, Phys.Rev.-Adv. Phys. {bf 44},405 (1995).\\
\noindent [3] C.W.J.Beenakker, Phys.Rev.Lett.{\bf 64}, 216 (1990).\\
\noindent [4] H.S.Swim and K.J.Chang and G.Ihm, Phys.Rev.Lett.{\bf 82},596(1999).\\
\noindent [5] Leinaas and J. Myrheim, Nuov. Cim. Soc.Itl.Fis {\bf B37},1 (1977).\\
\noindent [6] D.Arovas, J.R. Schrieffer and F.Wilczek,Phys.Rev.Lett.
{\bf 53}, 722(1984)\\
\noindent [7] B.I.Halperin , Phys. Rev. Lett. {\bf 52},1583 (1984).\\
\noindent [8] J.K. Jain et.al. Phys.Rev.Lett {\bf 91}, 036801 (2003)\\
\noindent [9] F.Wilczek,{\it Fractional Statistics,Anon Superconductivity,
  World scientific,1990}. \\
\noindent [10] F.D.M.Haldane, Phys. Rev.Lett {\bf 67} (1991)937.\\
\noindent [11] D.Banerjee, Phys.Lett. {\bf 269} (2000) 38.\\
\noindent [12] D.Banerjee, Mod.Phys.Lett. {\bf B5} (2000)181.\\
\noindent [13] D.Banerjee, Phys.Rev.{\bf B58} (1998) 4666.\\
\noindent [14] C.L.Kane,cond-mat/0210621-Apr 2003.\\
\noindent [15] D.V.Averin and V.J.Goldman,{\it Quantum Computation with quasi
particles of the Fractional Quantum Hall Effect}.\\
\noindent [16] F.D.M.Haldane,Phys.Rev.Lett.{\bf 51}, 605 (1983).  \\
\noindent [17] A. Lopez and Eduaro Fradkin, cond-mat/0310128, Dec 2003.\\
\noindent [18] U.Zullicke, J.J. Palacios and A.H. MacDonald,cond-mat/0301079, 2003.\\
\noindent [19] S.S.Mandal and J.K.Jain, Phys.Rev.{\bf B66}, 155302 (2002).\\
\noindent [20] S.Das Sarma, American Scientist {\bf 89},516 (2001).\\
www.americanscientist.org/articles/01articles/dassarma.html.\\

\end{document}